\documentclass[traditabstract]{aa}
\usepackage[T1]{fontenc} 
\usepackage{subfigure}
\usepackage{graphicx}

\begin{document}

\title{Launching of jets by cold, magnetized disks in Kerr Metric}

   \author{
              Aleksander S\k{a}dowski\inst{1}
          \and
              Marek Sikora \inst{1}
          }
   \institute{
             N. Copernicus Astronomical Center, Polish Academy
             of Sciences,
             Bartycka 18, 00-716 Warszawa,
             Poland \\
             \email{as@camk.edu.pl}
             \email{sikora@camk.edu.pl}        
             }
\date{Received ????; accepted ???? }
\abstract{
We confirm  discovery by Cao that in the vicinity of fast 
rotating black holes jets can be launched centrifugally by cold, 
magnetized disks even for nearly vertically shaped  magnetic flux surfaces.
Outflows produced under such extreme conditions are investigated  via
studying kinematics of test particles in the force-free magnetosphere
approximation. Implications of a possibility of magneto-centrifugal
launching of very well collimated central outflows around 
the fast rotating black holes are discussed in the general context of  
the jet formation scenarios in AGNs.}


\authorrunning{A. S\k{a}dowski, M. Sikora}
\titlerunning{Launching of jets}
  \maketitle

\section{Introduction}
According to a  popular class of astrophysical jet models they are powered
by rotational energy of accretion disks and 
mass loaded via 'magnetocentrifugal' forces. 
For Newtonian accretion disks such models
require the poloidal magnetic field lines to be inclined to the 
geometrically thin accretion disk by less than 60 degrees,
independently on a distance from the central object
(Blandford \& Payne 1982). For disks around  
black holes (BHs) this angle depends on the radius 
and the BH spin (Cao 1997; Lyutikov 2009). It is the larger the smaller is 
the distance  and the larger is the BH spin. At the inner edge of 
a Keplerian disk (located at the marginally stable orbit - $r_{ms}$) 
and for the maximal BH spin it approaches $90^{\circ}$.
We confirm these results and investigate the 
kinematics of test particles in the fixed force-free magnetosphere.  
Our paper is organized as follows. General equations 
describing  the particle kinematics are derived in \S2.
Geometry of the effective equipotential surfaces and of 
``light cylinders'' are presented in \S3.
Kinematics of test particles and its dependence on the BH spin 
and the launching distance is illustrated in \S4. Results are discussed 
in a general context of production of relativistic 
jets in active galactic nuclei in \S5 and summarised in \S6. 
 
\section{A rigid rotation of particle trajectories in the Kerr metric}

Kinematics of particles 
forced by magnetic fields to move on rigidly rotating
trajectories is convenient to investigate in a frame co-rotating
with magnetic field lines.
In such a frame the norm of a 4-velocity, $u_iu^i=-1$, is in the Kerr metric 
given by equation: 
\begin{eqnarray}
(g_{tt} + 2 g_{t\phi} \Omega_0 + g_{\phi\phi} \Omega_0^2) (u^t)^2  + 
2 (g_{t\phi} + g_{\phi\phi}\Omega_0) u^t u^{\phi'} &+&\\\nonumber
 g_{\phi\phi} (u^{\phi'})^2
+ g_{rr} (u^r)^2 + g_{\theta\theta} (u^{\theta})^2 &=& -1\, , 
\end{eqnarray}
where
$\phi' = \phi - \Omega_0 t$, $u^{\phi'} = u^{\phi} - \Omega_0 u^t$,
$\Omega_0 =$ const, the $t$, $\phi$, $r$, and $\theta$ 
are the Boyer-Lindquist coordinates, and
$g_{ik} = g_{ik}(r,\theta; a)$ are the Kerr metric components.
Through the paper the following quantities are expressed in dimensionless
units: 
radius $r: r/(GM/c^2)$, BH spin $a^*= J/(GM^2/c)$, and angular velocity
$\Omega_0: \Omega_0/(c^3/GM)$, where $M$ and $J$ are the BH mass and 
angular momentum, respectively. 

For a particle trajectory which in the rotating frame is described by
functions $r = r(\theta)$ and $\phi' = \phi'(\theta)$, the $r$ and 
$\phi'$ components of the particle 4-velocity are
$u^{r} = r_{,\theta} u^{\theta}$ and 
$u^{\phi'} = {\phi'}_{,\theta} u^{\theta}$, respectively. Then the norm 
of the 4-velocity reads
\begin{equation}
\tilde g_{tt} (u^t)^2 +2\tilde g_{t\theta} u^tu^{\theta} +\tilde g_{\theta\theta}
(u^{\theta})^2 = -1 \,,
\end{equation}
where
\begin{equation}
\tilde g_{tt} = g_{tt} + 2g_{t\phi}\Omega_0 + g_{\phi\phi}\Omega_0^2 \, , 
\end{equation}
\begin{equation}
\tilde g_{t\theta} = (g_{t\phi} + g_{\phi\phi} \Omega_0) {\phi'}_{,\theta} \, ,
\end{equation}
\begin{equation}
\tilde g_{\theta\theta} = g_{\theta\theta} + g_{rr} (r_{,\theta})^2 + 
g_{\phi\phi} ({\phi'}_{,\theta})^2 
\, . 
\end{equation}
Noting that 
\begin{equation}
\tilde u_t  = \tilde g_{tt} u^t + \tilde g_{t\theta} u^{\theta} = 
-\tilde \epsilon \, ,
\end{equation}
where $\tilde \epsilon$ is the constant of motion of particle moving
along a rigidly rotating trajectory, one can find inserting $u_t$ from
Eq. (6) into Eq. (2) that
\begin{equation}
u^{\theta} = 
\sqrt{{(\tilde \epsilon)^2 - (-\tilde g_{tt}) \over 
(-\tilde g_{tt})\tilde g_{\theta\theta} + (\tilde g_{t\theta})^2}}\, ,
\end{equation}
and 
\begin{eqnarray}
u^t &=& {\tilde \epsilon \over (-\tilde g_{tt})} + 
{\tilde g_{t\theta} \over (-\tilde g_{tt})} \, u^{\theta} =\\\nonumber
&=&{\tilde \epsilon \over (-\tilde g_{tt})} + 
{\tilde g_{t\theta} \over (-\tilde g_{tt})} \,
\sqrt{{(\tilde \epsilon)^2 - (-\tilde g_{tt}) \over 
(-\tilde g_{tt})\tilde g_{\theta\theta} + (\tilde g_{t\theta})^2}} \, .
\end{eqnarray}

Hence, for a fixed particle trajectory and a given constant of motion 
$\tilde \epsilon$ 
kinematics of a test particle is fully 
determined. This kinematics can be illustrated in the locally non-rotating 
frame (Bardeen, Press, \& Teukolsky 1972), i.e. the frame 
of zero-angular-momentum-observers (ZAMO). In such a frame the line 
element is 
\begin{equation}
ds^2 = (g_{tt} + \omega g_{t\phi}) dt^2 + g_{\phi\phi}(d\phi - \omega dt)^2
+ g_{rr}dr^2 + g_{\theta\theta} d\theta^2 \, ,
\end{equation}
where $\omega= -g_{t\phi}/g_{\phi\phi}$ is the angular velocity of 
'the dragged inertial frames'. 
Then the projection of the 4-velocity onto the orthonormal tetrad of the local
Minkowski space can be used to calculate  the Lorentz factor  and velocity
components of a particle in the ZAMO frame:
\begin{equation}
\gamma = u^{(t)} = u^t \,
\sqrt{-g_{tt} + g_{t\phi}^2/g_{\phi\phi}} \, ,
\end{equation}
\begin{equation}
v^{(\phi)} = {u^{(\phi)} \over u^{(t)}} =
\left({{\phi'}_{,\theta} u^{\theta} \over u^t} + (\Omega_0-\omega)\right)
\, \sqrt{g_{\phi\phi} \over -g_{tt} + g_{t\phi}^2/g_{\phi\phi}} \, ,
\end{equation}
\begin{equation}
v^{(p)} \equiv \sqrt{{v^{(r)}}^2 + {v^{(\theta)}}^2} =
{u^{\theta} \over u^t} \,  \sqrt{{g_{rr} r_{,\theta}+g_{\theta\theta}
\over -g_{tt} + g_{t\phi}^2/g_{\phi\phi}}} \, .
\end{equation}
where the physical velocities $v^{(i)}$ are expressed in the speed of light
units.

\section{Effective potential}

\begin{figure}
\centering
 \subfigure[$a^*=0$]
{
\includegraphics[width=.45\textwidth]{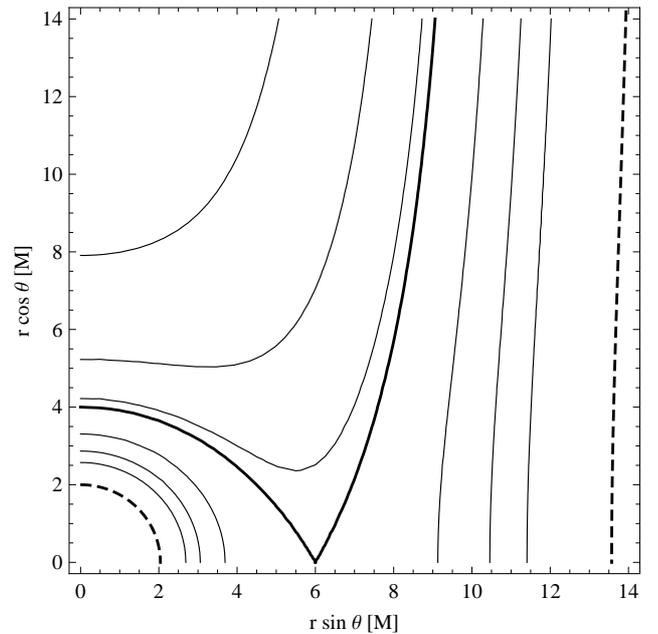}
}
\\
\subfigure[$a^*=0.99$]
{
\includegraphics[width=.45\textwidth]{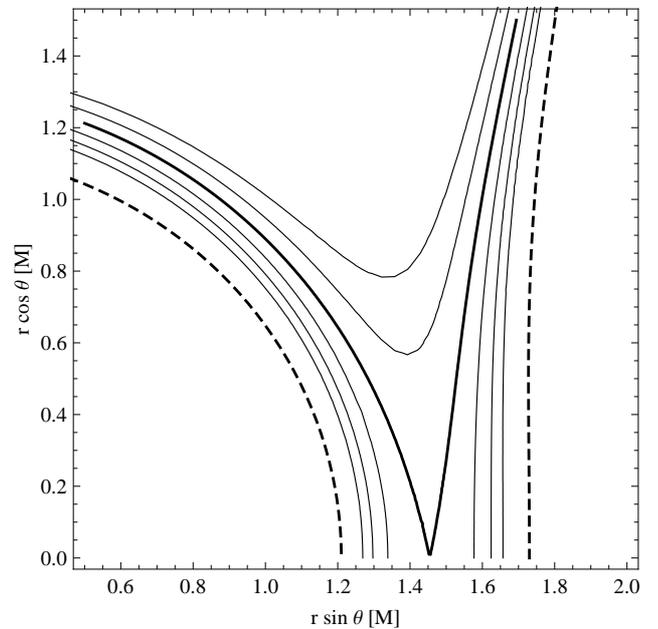}
}
\caption{Equipotential surfaces of the effective potential 
($V_{eff} = \sqrt{-\tilde g_{tt}}$) for $\Omega_0=\Omega_0(r=r_{ms},a^*)$ for 
a non-rotating (top) and spinning (bottom panel) BHs. The thick solid lines represent $V_{eff}(r,\theta)=V_0\equiv V_{eff}(r=r_{ms},\theta=\pi/2)$ being the effective potential crossing the equator at $r=r_{ms}$. The dashed lines present locations of the inner and outer light cylinders. The thin solid lines are drawn for the following values of the effective potential: $\frac69$, $\frac79$, $\frac89$, $\frac{10}9$, $\frac{11}9V_{0}$.}
\label{fig:1}
\end{figure}

\begin{figure}
\centering
\includegraphics[width=.45\textwidth]{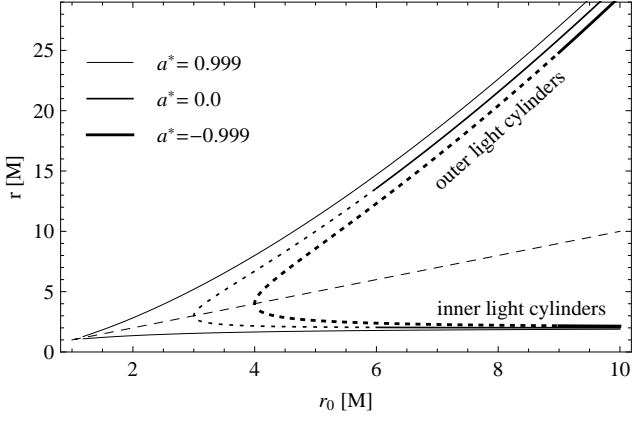}
\caption{Locations of the inner and outer light cylinders for $\theta=\pi/2$ as a function of radius $r_0$ defining the angular velocity $\Omega_0$. Inner and outer cylinders are represented by curves below and above the dashed line $r=r_0$, respectively. Dotted curves denote locations of light cylinders for $r_0<r_{ms}$. Profiles for three values of BH spin: $-0.999$, $0$, $0.999$ are presented.}
\label{fig:2}
\end{figure}

\begin{figure}
\centering
\includegraphics[width=.45\textwidth]{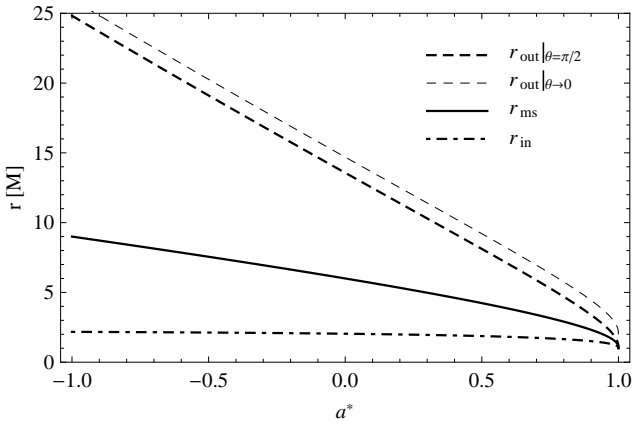}
\caption{Locations of the light cylinders for $\Omega_0=\Omega_0(r=r_{ms})$ versus BH spin. Location of the outer light cylinder is shown for the equatorial plane (thick, dashed line) as well as its asymptotic ($\theta\rightarrow0)$ value (thin, dashed line).}
\label{fig:3}
\end{figure}

An effective potential defined as the minimum energy of test particles
forced to rotate with a given angular velocity $\Omega_0$ is   
$V_{eff} = \sqrt{-\tilde g_{tt}}$ (obtained from Eq.~(7) setting $u^\theta=0$). Its  equipotential surfaces, 
$V_{eff}(r,\theta)=const$, are illustrated in Fig.~\ref{fig:1}. They are enclosed 
between the inner and outer `light cylinders' given by $\tilde g_{tt} = 0$ 
(Lyutikov 2009). Locations of the inner and outer 
light cylinders in the equatorial plane are presented in Fig.~\ref{fig:2}. 
For a given spin the cylinders coincide at the photon orbit. For $a^*>0.91$ and the angular velocity $\Omega_0$
corresponding to the marginally stable orbit $r_{ms}$ the outer light cylinder is 
enclosed by the BH ergosphere. In Fig.~\ref{fig:3} we show the dependence of 
the equatorial and asymptotic locations of 
the outer light cylinder on the BH spin for $\Omega_0$ calculated at 
the marginally stable orbit. The asymptotic radius is always close to its equatorial 
plane value (e.g. $14.7M$ vs $13.6M$ for a non-rotating BH).

\begin{figure}
\centering
\includegraphics[width=.45\textwidth]{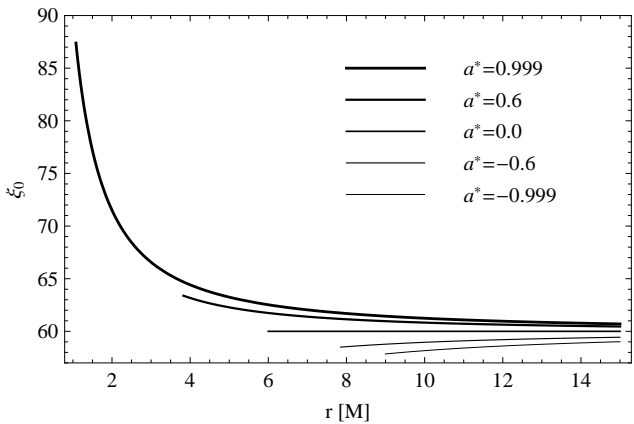}
\caption{The critical angle $\xi_0$ dependence on the radius $r_0$. Profiles for different BH spins, starting at the marginally stable orbits, are shown.}
\label{fig:4}
\end{figure}

Equipotential surfaces, $dV_{eff}=0$, intersect the equatorial plane at angles 
%
\begin{equation}
\tan \xi = -r \left(d\theta \over dr \right)_{\theta=\pi/2} =
 r \left( V_{eff,r} \over V_{eff,\theta} \right)_{\theta = \pi/2} \, .
\end{equation}
Since $V_{eff,\theta}|_{\theta=\pi/2} = 0$, while $V_{eff,r}|_{\theta=\pi/2} = 0$
only at $r=r_0$ at which the Keplerian law is satisfied, i.e. 
\begin{equation}
\Omega_0 = {1 \over r_0^{3/2} +a} \, .
\end{equation}
this angle is $\xi = \pi/2$ for all radii but $r=r_0$. 
At $r=r_0$ this angle can be found  by applying 
in Eq. (13) the L'Hospital's rule and by noting that
$V_{eff,r\theta}|_{\theta=\pi/2}=0$. This gives
\begin{eqnarray}
\tan{\xi_0} &=& 
r_0 \sqrt{-\left(V_{eff,rr} \over V_{eff,\theta\theta} \right)_{\theta=\pi/2, r=r_0}}=\\\nonumber
&=&\sqrt{3 \over 1-4a r_0^{-3/2} + 3a^2 r_0^{-2}} \, .
\end{eqnarray}
This formula fully agrees with the formula obtained by Cao (1997) and 
Lyutikov (2009).
Dependence of $\xi_0$ on the radius $r_0$
and on the BH spin is shown in Fig.~\ref{fig:4}.

\section{Kinematics of test particles in a force-free magnetosphere}

\subsection{Purely poloidal magnetic fields}
For a strong  dynamical domination of large scale magnetic fields driven
by an accretion disk, 
test particles are restricted to move along magnetic field lines. Being 
anchored to
a differentially rotating accretion disk, the poloidal magnetic field
generates differentially rotating magnetosphere.  Trajectories of test
particles can then be identified with magnetic field lines rotating
with an angular velocity of their foot-points (Ferraro 1937).
For purely poloidal magnetic fields such trajectories are in the
co-rotating frame planar ($\phi'=const$), the metric component 
$\tilde g_{t\theta} = 0$ and equations (7) and (8) 
reduce to
\begin{equation}
u^{\theta} = \sqrt{ {(\tilde \epsilon_0)^2 - (-\tilde g_{tt}) \over
(-\tilde g_{tt}) \tilde g_{\theta\theta}}} \, ,
\end{equation}
\begin{equation}
u^t= {\tilde \epsilon_0 \over (-\tilde g_{tt})} \, ,
\end{equation}
where $\tilde \epsilon_0^2 = - \tilde g_{tt} (\theta=\pi/2, r=r_0)$,
$\tilde g_{ik} = \tilde g_{ik}(\theta, r(\theta; r_0))$, and 
$r(\theta; r_0)$ is the shape of the magnetic field surface which is 
determined by the shape of poloidal magnetic field lines. 

Test particles can be pulled  from the cold disk by centrifugal forces 
provided the inclination of magnetic field surfaces  
at the foot-point is smaller than $\xi_0$ and they 
can proceed further provided the shape of poloidal magnetic field lines  
is such that along them the condition $-\tilde g_{tt} < \tilde \epsilon_0^2$
is satisfied.

Examples of kinematics of the test particles are plotted in Fig. 4.
The illustrated cases are calculated for magnetic field lines
anchored at the marginally stable orbit and having shape
\begin{equation}
r = r_0 {\tan{\xi_0} \over (\sin{\theta} \tan{\xi} - \cos{\theta})} \, ,
\end{equation}
which in the coordinate plane ($r\cos{\theta}$, $r\sin{\theta}$)
is the straight line anchored at $r=r_0$ and inclined at the angle $\xi$.
The presented velocities and Lorentz factors were calculated in the locally
non-rotating frame (see Eqs.~(10)-(12)).
As we can see from Eq.~(8), the particles which are forced to
move on rigidly rotating planar trajectories approach the speed of light
at the light cylinder. 
This simply demonstrates that no physical solutions
exists on and beyond light cylinders for purely poloidal magnetic field 
structures.  

\begin{figure}
\centering
\includegraphics[width=.45\textwidth]{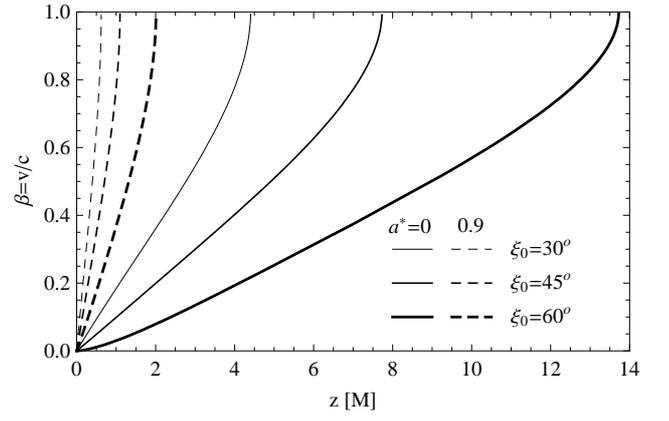}
\caption{Dynamics of a particle moving along a purely poloidal magnetic field lines achored at $r_{ms}$ and inclined at different angles for non-spinning (solid) and spinning (dashed lines) BHs. The horizontal axis represents the vertical coordinate $z$.}
\label{fig:5}
\end{figure}

\subsection{Force-free magnetosphere with non-zero azimuthal magnetic field
component}

Not necessarily such constraints apply for magnetosphere if taking into account
presence of toroidal magnetic field component which is induced by poloidal
electrical currents. In this case the Ferraro's law about iso-rotation
still applies, i.e.  magnetic field lines
and, therefore, particle trajectories, still rotate rigidly, however, due to 
the presence
of the toroidal magnetic component the rotating  trajectories are not planar
and the test particles may pass the light cylinder sliding on magnetic field 
lines in the opposite direction to the rotation of magnetic field surfaces,
i.e. with $d\phi'/dt <0$,
which takes place if ${\phi'}_{,\theta} <0$. 
This can be concluded from Eq.~(8) after rewriting it in the form
\begin{equation}
u^t = {\tilde \epsilon_0 \over (-\tilde g_{tt})} 
\left(1-
\sqrt{{1-(-\tilde g_{tt})/\tilde \epsilon_0^2 \over 1 + (-\tilde g_{tt}) 
\tilde g_{\theta\theta} /\tilde g_{t\theta}^2}}
\right)  \, ,
\end{equation}
where the metric functions $\tilde g_{ik}$ are calculated along 
trajectories $r=r(\theta; r_0)$ and $\phi'= \phi'(\theta; r_0)$.
At the light cylinder both denominator and numerator 
($-\tilde g_{tt}$ and the expresion in brackets)
become zero and
applying the L'Hospital's rule one can check that $u^t$ and, therefore,
$\gamma$ given by Eq.~(10) is finite. Hence, the test particle can pass the 
light cylinder and proceed further provided 
\begin{equation}
(\tilde g_{t\theta})^2 > \tilde g_{tt} \, \tilde g_{\theta\theta} \, ,
\end{equation}
i.e. for respectively large negative values of ${\phi'}_{,\theta}$,
\begin{equation}
({\phi'}_{,\theta})^2 > {\tilde g_{tt} (g_{\theta\theta} + g_{rr}(r_{,\theta})^2)
\over (g_{t\phi}+g_{\phi\phi}\Omega_0)^2 - g_{\phi\phi} \tilde g_{tt}} \, .
\end{equation}

\begin{figure}
\centering
 \subfigure[$a^*=0,\quad\xi=60^o$]
{
\includegraphics[width=.45\textwidth]{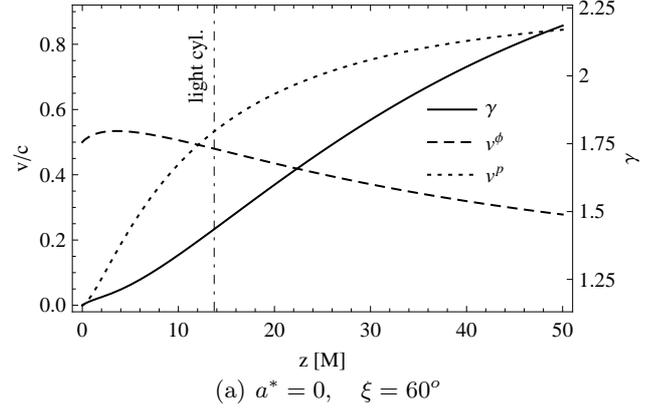}
}
\\
\subfigure[$a^*=0.99,\quad\xi=78.14^o$]
{
\includegraphics[width=.45\textwidth]{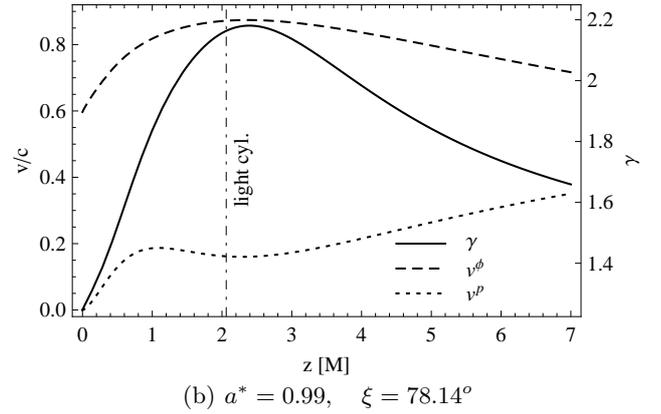}
}
\caption {The velocity components and the Lorentz factor $\gamma$ of particles moving along magnetic field lines with non-zero azimuthal component anchored at $r_{ms}$ and inclined at the critical angle $\xi=\xi_0$ for non-spinning (top) and spinning (bottom panel) BHs. Poloidal and azimuthal velocity components are marked by dotted and dashed lines, respectively. Solid curves present the Lorentz factor profiles.}
\label{fig:6}
\end{figure}

\begin{figure}
\centering
\includegraphics[width=.45\textwidth]{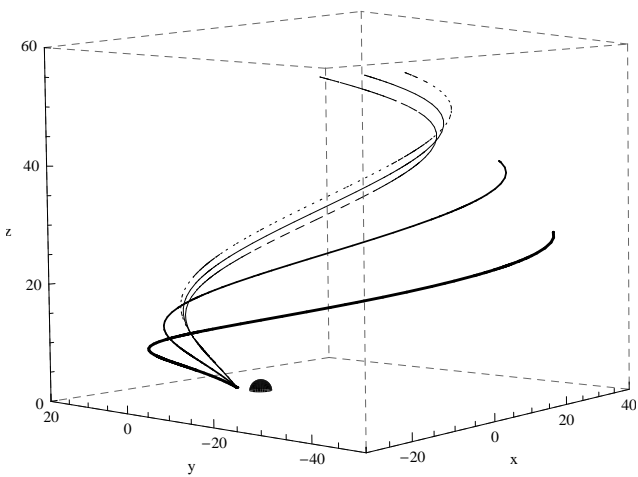}
\caption{Shape of the magnetic field lines with $\eta=1.0$ for $\xi=30^o$ (the thickest solid line), $\xi=45^o$ and $\xi=60^o$ (the thinnest solid line). For the latter ($\xi=60^o$) two other curves are presented for $\eta=0.95$ (dotted line) and $\eta=1.05$ (dashed line). All lines are anchored at $r_{ms}$ and the BH has zero spin. The black spot denotes location of the BH horizon.}
\label{fig:8}
\end{figure}

Examples of kinematics of test particles for helical magnetic field 
structure  are 
illustrated in Fig.~\ref{fig:6}. The velocity components and the Lorentz factor are
computed for $r(\theta; r_0)$ given by Eq.~(18) and for
\begin{equation}
\phi' = -{\eta \Omega_0 \over \cos{\xi}}\, (r \sin{\theta} - r_0) \, .
\end{equation}
%
The above choice of  $\phi'(\theta; r_0)$ corresponds roughly with
a  poloidal magnetic field component scaling as $B_p \sim 1/(r \sin{\theta})^2$
and toroidal component scaling as $B_{\phi} \sim 1/( r \sin{\theta})$, while
$\eta \approx -B_p/B_{\phi}|_{r\sin{\theta}=1/\Omega_0}$ (in Fig.~\ref{fig:8} we plot
a visualisation of such a shape of the magnetic field lines). 

The upper panel
of Fig.~\ref{fig:6} presents the velocity and Lorentz factor profiles for a non-spinning
BH. The particle is initially at rest with respect to the disk --- its poloidal 
velocity component in the ZAMO frame is zero while its azimuthal velocity corresponds
to the Keplerian angular velocity at $r_0$. Once the particle leaves the equilibrium
it starts to be centrifugally accelerated along the magnetic field line. The poloidal
velocity component rapidly increases. Due to the fact that the magnetic field lines
are inclined also in the azimuthal direction the particle starts to slide along them
in the direction opposite to the disk rotation. Therefore, its angular velocity 
decreases below the initial value $\Omega_0$. The corresponding profile of 
the physical
velocity in the azimuthal direction $v^{(\phi)}$ is shown. The light cylinder
is crossed at $z\approx14\rm M$ with both velocity components close to 
$0.5\rm c$.
The profile of the Lorentz $\gamma$ factor is shown with the solid line 
(please note the 
vertical scale for $\gamma$ is marked on the right axis). Initially, 
it corresponds 
to the Keplerian angular velocity ar $r_{ms}$ ($\approx 1.2$ for $a^*=0$ case). At the light
cylinder $\gamma\approx 1.7$. 
The bottom panel of Fig.~\ref{fig:6}
presents a similar study of particle dynamics for a rotating BH case. Its behaviour
is similar with exception to the fact that the $\gamma$ factor is initially strongly
dominated by the azimuthal velocity component. 

\begin{figure}
\centering
 \subfigure
{
\includegraphics[width=.45\textwidth]{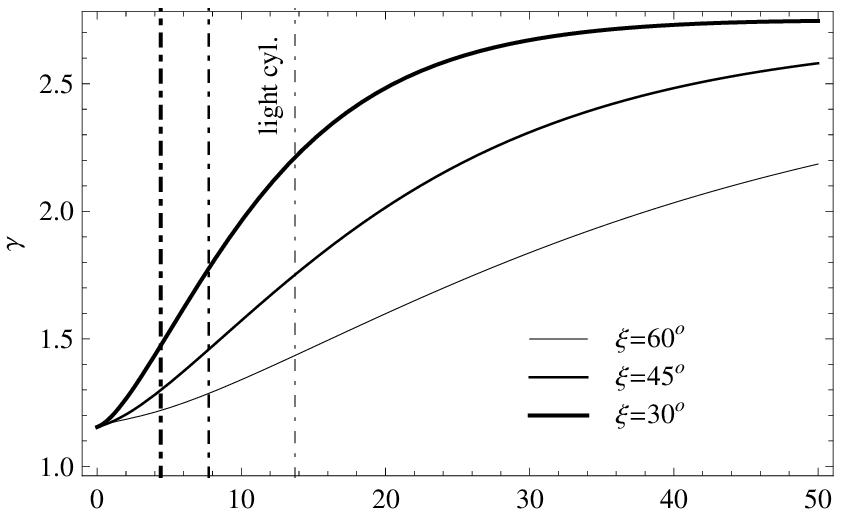}
}
\\
\vspace{-.04\textwidth}
 \subfigure
{
\includegraphics[width=.45\textwidth]{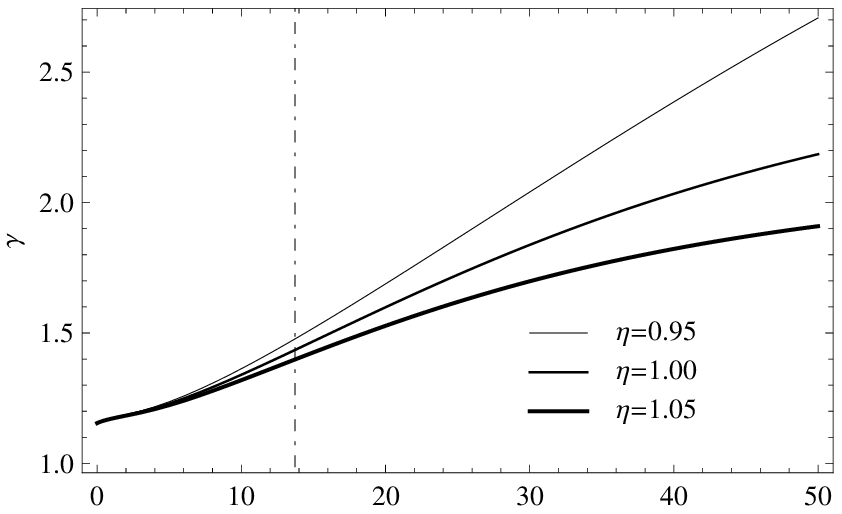}
}
\\
\vspace{-.04\textwidth}
 \subfigure
{
\includegraphics[width=.45\textwidth]{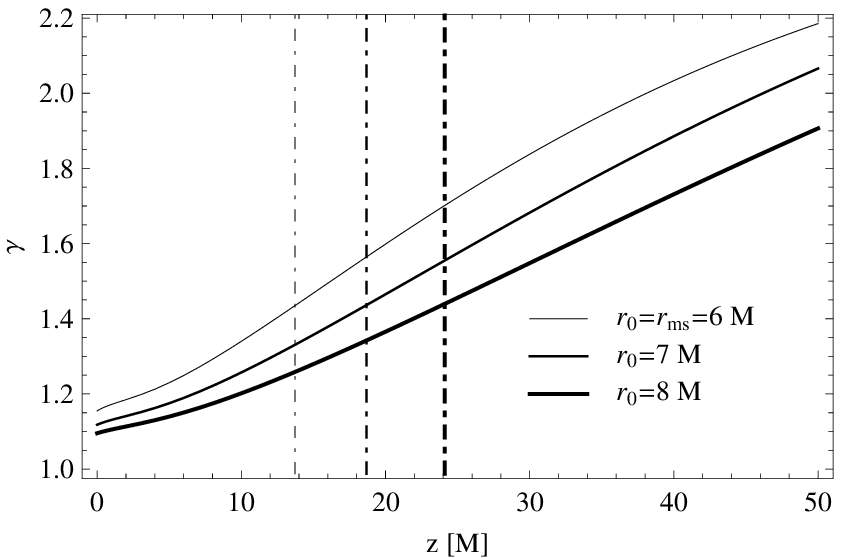}
}
\caption {The Lorentz $\gamma$ factor dependence on the magnetic field line inclination angle $\xi$ (top panel), $\eta$ parameter (middle panel) and the $r_0$ radius (bottom panel) for a non-spinning BH. The vertical dashed lines denote crossing through the outer light cylinder. If not stated otherwise the parameters are: $\xi=60^o$, $\eta=1.0$ and $r=r_{ms}$. }
\label{fig:7}
\end{figure}

We demonstrate in Fig.~\ref{fig:7} the dependence of the test particle kinematics on 
$\xi$, $r_0$ and $\eta$. The upper panel shows that the lower the inclination
angle $\xi$ the more rapid is the particle acceleration. The middle panel presents
the dependence on the $\eta$ parameter which determines the magnetic field line
torsion. The more 'twisted` is the line (corresponding to increasing $\eta$) the easier 
particle slides in the azimuthal direction and the resulting $\gamma$ factor
increases slower. Finally, the lower panel presents the dependence on the 
radius $r_0$ defining the location of the magnetic field line foot-point. 

Obviously, illustrated  in this section kinematics has only exemplary 
character and applies  to the region in which the structure of the outflow
is strongly dominated by the magneto-dynamics. For low mass loading rates,
such a force-free approximation can be used to study the matter kinematics
even far beyond the light cylinder. However, for high mass loading rates the 
force-free approximation can break down even in the sub-Alfv\'enic region,
deep within the light cylinder.

\section{Discussion}

Magnetocentrifugal scenario is applicable to the production of jets
in 'Newtonian' objects like YSOs (Young Stellar Objects). However,
it cannot explain the whole jet production process in BH accretion systems. This is because
at least in idealized, steady, axisymmetric models with an accretion disk 
treated only as the boundary surface, efficient centrifugal mass loading makes 
the outflows too heavy to be accelerated up to the observed relativistic velocities.
Favored mass loading scenarios for such objects involve pair production by
photon-photon interactions. For typical BH accretion systems the coronal
activities are enough powerful to support in the vicinity of the BH
density of pairs by many orders of magnitude larger than the Goldreich-Julian 
density (Phinney 1983; McKinney 2005), which allows to treat the outflows in
the ideal MHD approximation. By the same time, the rest mass-energy density 
of the pair plasma  is by 
several orders lower than energy density of magnetic fields required to power 
luminous large scale jets in radio-loud objects, and this implies formation of
relativistic, at least initially strongly Poynting flux dominated, outflows.

The e$^{+}$e$^{-}$ jets can be powered by both rotating BHs 
(Blandford \& Znajek 1977; Phinney 1983; Beskin 1997) and accretion disks  
(Blandford 1976; Lovelace 1976; Lovelace, Wang \& Sulkanen 1987; 
Ustyugova et al. 2000).
Such jets need external collimation to be effectively accelerated
and converted to the matter dominated ones 
(Begelman \& Li 1994; Narayan , McKinney, \& Farmer 2007; 
Komissarov, et al. 2007; Lyubarsky 2009; Porth \& Fendt 2010). 
The collimation  is often considered to be provided
by slower, barionic MHD outflows launched in the accretion disk
(Sol, Pelletier, \& Asseo 1989; Bogovalov \& Tsinganos 2005; 
Gracia et al. 2005; Beskin \& Nokhrina 2006).
However, such a jet structure may need modification for jets launched 
around fast rotating BHs if the possibility of centrifugal launching
of proton-electron outflows from innermost portions 
of accretion disk was taken into account.
Since the inclination angle of the effective potential surfaces is 
maximal at the inner edge of a disk and rapidly drops with a radius
(see Eq.~15 and Fig.~\ref{fig:4}, the p-e dominated 
outflow from the innermost portions of accretion disk is embraced by 
the Poynting flux dominated electron-positron outflow. 
This implies less 
relativistic spines and more relativistic sheaths  of  jets produced by 
the accretion 
disk around fast rotating BHs, or a respective triple jet structure 
if taking into account the central contribution to the outflow from 
the BH magnetosphere.

The production of powerful jets by accretion disks, both centrifugally 
loaded and by pairs, requires strong large scale magnetic fields. 
Possibility  of dragging of such fields by accreting plasma,
originally suggested by Bisnovatyi-Kogan \& Ruzmaikin (1974), is still 
debated (Lubow, Papaloizou, \& Pringle 1994; Spruit \& Uzdensky 2005;
Bisnovatyi-Kogan \& Lovelace 2007; Rothstein \& Lovelace 2008;
Beckwith, Hawley, \& Krolik 2009). Related to the lack of the profound model 
of the evolution of magnetic fields in accretion disks, unknown remains
radial distribution of inclination of magnetic field lines which
is critical to establish a launching distance domain of centrifugal outflows
and their initial collimation. And finally, even for a fixed 
large scale magnetic field intensity and geometry, mass loading rate
and therefore the terminal speed of the centrifugal outflow depends very much
on details of the vertical disk structure (Ogilvie \& Livio 2001), which
due to severe uncertainties  is usually ignored.
In particular, due to partial losses 
of angular momentum (taken away by the outflow) the boundary layers of 
the accretion disk likely become sub-Keplerian. As a result, proton-electron
outflow may need some initial boost to pass the created potential barrier.
It can be provided by heating or mechanically  by some magnetic flaring 
activities and/or by radiation pressure of effectively super-Eddington  
flux.   Existence of the additional potential barrier may significantly
limit mass loading rate allowing
proton-electron loaded outflows to reach at least mildly relativistic 
speeds.
The resulting jet structure --- mildly relativistic proton-electron component
sandwiched between the pair dominated relativistic spine and sheath ---
albeit very speculative at the moment, is very 
promising from the observational point of view because may explain 
a significant  proton content of AGN jets deduced from analyses of 
a matter content in blazars (Sikora \& Madejski 2000; Ghisellini et al. 2009) 
and in radio lobes of 
powerful radio galaxies (Stawarz et al. 2007; Perlman et al. 2009). 
 
\section{Conclusions}

Main results of this paper can be summarized as follows:

\noindent
- An effective potential and light cylinders are investigated in the rigidly
rotating frame in the Kerr metric. The intersection of equipotential surfaces 
with a geometrically thin disk at the annulus where a given angular velocity is
equal to the Keplerian velocity gives the critical angle below which 
a cold outflow can be launched by centrifugal forces. The location
of outer light cylinders is shown to depend strongly on radius and BH spin.
For BH spin $a^* > 0.91$ 
and the Keplerian angular velocity at the marginally stable orbit 
the light-cylinder radius in the equatorial plane
is enclosed within the erghosphere.

\noindent 
- The condition for magnetocentrifugal launching of jets obtained for
  Keplerian disks rotating around Kerr BHs by Cao (1997) and
Lyutikov (2009)
is confirmed. It tells us that the maximum  inclination angle
of magnetic flux surfaces at which cold matter can be extracted from the disk
depends on a distance from the BH and on the BH spin.
The condition shows that in case of very fast rotating BHs the central outflows 
can be launched  even along of almost vertically shaped  magnetic surfaces.
 
\noindent
-  We show how kinematics of test particles pulled by centrifugal forces from 
a Keplerian disk can be algebraically determined for a given magnetic field
structure in the force-free outflow approximation. Examples of 
test particle kinematics are illustrated 
and the condition for the toroidal magnetic field component is derived
to allow the  particle to cross  the light cylinder.  

\noindent
- Possible implications for a jet structure are discussed as imposed
by the condition for magnetocentrifugal launching of jets by inner portions 
of magnetized  disks around fast rotating BHs. In this case
a triple-component structure of a jet can be envisaged, with 
a proton-electron component of a jet being sandwiched between the
relativistic pair dominated spine and sheath.

\begin{acknowledgements} 
AS and MS acknowledge direct support by
Polish Ph.D. NN203 304035 and MNiSW NN203 301635
grants, respectively.
\end{acknowledgements} 



\begin{thebibliography}{}

\bibitem[Bardeen et al.(1972)]{bpt72}
Bardeen, J.M., Press, W.H., \& Teukolsky, S.A. 1972, ApJ, 178, 347

\bibitem[Beckwith et al.(2009)]{bhk09}
Beckwith, K., Hawley, J., \& Krolik, J.H. 2009, arXiv:0906.2784

\bibitem[Begelman \& Li(1994)]{BL94}
Begelman, M.C., \& Li, Z.-Y. 1994, ApJ, 426, 269

\bibitem[Beskin(1997)]{beskin97}
Beskin, V.S. 1997, Soviet Physics Uspekhi, 40, 659

\bibitem[Beskin \& Nokhrina(2006)]{BN06}
Beskin, V.S., \& Nokhrina, E.E. 2006, MNRAS, 367, 375

\bibitem[Bisnovatyi-Kogan \& Lovelace(2007)]{BL07}
Bisnovatyi-Kogan, G.S., \& Lovelace, R.V.E. 2007, ApJ, 667, L167

\bibitem[Bisnovatyi-Kogan \& Ruzmaikin(1974)]{BR74}
Bisnovatyi-Kogan, G.S., \& Ruzmaikin, A.A. 1974, Ap\&SS, 28, 45

\bibitem[Blandford(1976)]{Bland76}
Blandford, R.D. 1976, MNRAS, 176, 465

\bibitem[Blandford \& Payne(1982)]{bp82}
Blandford, R.D. \& Payne, D. G. 1982, MNRAS, 199, 883

\bibitem[Blandford \& Znajek(1977)]{BZ77}
Blandford, R.D., \& Znajek, R.L. 1977, MNRAS, 179, 433

\bibitem[Bogovalov \& Tsinganos(2005)]{BT05}
Bogovalov, S.V., \& Tsinganos, K. 2005, MNRAS, 357, 918

\bibitem[Cao(1997)]{cao97}
Cao, X. 1997, \mnras, 291, 145

\bibitem[Ferraro(1937)]{1937MNRAS..97..458F} Ferraro, V.~C.~A.\ 1937, 
\mnras, 97, 458 

\bibitem[Ghisellini et al.(2009)]{gtfgmc09}
Ghisellini, G., Tavecchio, F., Foschini, L., Ghirlanda, G., Maraschi, L.,
\& Celotti, A. 2009, arXiv:0909.0932

\bibitem[Gracia et al.(2005)]{gtg05}
Gracia, J., Tsinganos, K., \& Bogovalov, S.V. 2005, A\&A, 442, L7

\bibitem[Komissarov et al.(2007)]{kbvk07}
Komissarov, S.S., Barkov, M.V., Vlahakis, N., \& K\"onigl, A. 2007, MNRAS, 380,
51

\bibitem[Lovelace(1976)]{Lov76}
Lovelace, R.V.E. 1976, Nat, 262, 649

\bibitem[Lovelace et al.(1987)]{LWS87}
Lovelace, R.V.E., Wang, J.C.L., \& Sulkanen, M.E. 1987, ApJ, 315, 504

\bibitem[Lubow et al.(1994)]{lpp94}
Lubow, S.H., Papaloizou, J.C.B., \& Pringle, J.E. 1994, MNRAS, 267, 235

\bibitem[Lyubarsky(2009)]{Lyu09}
Lyubarsky, Y.E. 2009, MNRAS, in press, arXiv:0909.4819

\bibitem[Lyutikov(2009)]{L09}
Lyutikov, M. 2009, MNRAS, 396, 1545 

\bibitem[McKinney(2005)]{McK05}
McKinney, J.C. 2005, arXiv:astro-ph/0506368

\bibitem[Narayan et al.(2007)]{nkf07}
Narayan, R., McKinney, J., \& Farmer, A. 2007, MNRAS, 375, 548

\bibitem[Ogilvie \& Livio(2001)]{OL01}
Ogilvie, G.I., \& Livio, M. 2001, ApJ, 553, 158

\bibitem[Perlman et al.(2009)]{pgmk09}
Perlman, E.S., Georganopoulos, M., May, E.M., \& Kazanas, D. 
2009, arXiv:0910.3021

\bibitem[Phinney(1983)]{Ph83}
Phinney, E.S. 1983, PhD thesis, Cambridge University

\bibitem[Porth \& Fendt(2009)]{PF09}
Porth, O., \& Fendt, C. 2009, arXiv:0911.3001

\bibitem[Rothstein \& Lovelace(2008)]{RL08}
Rothstein, D.M., \& Lovelace, R.V.E. 2008, ApJ, 677, 1221

\bibitem[Sikora \& Madejski(2000)]{SM00}
Sikora, M., \& Madejski, G. 2000, ApJ, 534, 109

\bibitem[Sol et al.(1989)]{spa89}
Sol, H., Pelletier, G., \& Asseo, E. 1989, MNRAS, 237, 411

\bibitem[Spruit \& Uzdensky(2005)]{SU05}
Spruit, H.C., \& Uzdensky, D.A. 2005, ApJ, 629, 960

\bibitem[Stawarz et al.(2007)]{scho07}
Stawarz, {\L}., Cheung, C.C., Harris, D.E., \& Ostrowski, M.
2007, ApJ, 662, 213

\bibitem[Ustyugova et al.(2000)]{urllc00}
Ustyugova, G.V., Lovelace, R.V.E., Romanova, M.M., Li, H., \& Colgate, S.A.
2000, ApJ, 541, L21

\end{thebibliography}
\end{document}